\documentclass[letter]{aa} 
\usepackage{graphicx}
\usepackage{txfonts}
\begin{document}
   \title{Trigonometric parallaxes of ten ultracool subdwarfs\thanks{Based on observations collected with Omega 2000 at the 3.5m telescope at the Centro Astron{\'o}mico Hispano Alem{\'a}n (CAHA) at Calar Alto, operated jointly by the Max-Planck Institut f\"ur Astronomie and the Instituto de Astrof{\'i}sica de Andaluc{\'i}a (CSIC))}}


   \author{E. Schilbach\inst{1}\and
           S. R\"oser\inst{1}\and
           R.-D. Scholz\inst{2}
          }
   \institute{Astronomisches Rechen-Institut, 
              Zentrum f\"ur Astronomie der Universit\"at Heidelberg, 
              M\"onchhofstra{\ss}e 12-14, 69120 Heidelberg, Germany\\
              \email{[elena,roeser]@ari.uni-heidelberg.de}
          \and
              Astrophysikalisches Institut Potsdam,
              An der Sternwarte 16, 14482 Potsdam, Germany\\
              \email{rdscholz@aip.de}
             }

   \date{Received 4 November 2008; accepted 25 November 2008}


  \abstract
   {}
   {We measure absolute trigonometric parallaxes and proper motions with respect to 
many background galaxies for a sample of ten ultracool subdwarfs.}
   {Observations were taken in the $H$-band with the OMEGA2000 camera at the
3.5m-telescope on Calar Alto, Spain during a
time period of 3.5 years. For the first time,  the reduction of the astrometric
measurements was carried out directly with respect to background galaxies.
We obtained absolute parallaxes with mean errors
ranging between 1 and 3 mas.
   }
   {
With six completely new parallaxes we more than doubled the
number of benchmark ultracool ($>$sdM7) subdwarfs. 
Six stars in the $M_{K_s}$ vs. $J-K_s$ diagram fit perfectly
to model subdwarf sequences from M7 to L4 with $[M/H]$ between 
$-$1.0 and $-$1.5, whereas 4  are consistent with a moderately
low metallicity ($[M/H]=-0.5$) from M7 to T6.
All but one of our objects have large tangential velocities between
200 and 320 km/s typical of the Galactic halo population.

Our results are in good agreement with recent independent measurements
for three of our targets and confirm the previously measured parallax
and absolute magnitude $M_{K_s}$ of the nearest and coolest (T-type) subdwarf
2MASS 0937+29 with higher accuracy.

For all targets, we also obtained infrared $J,H,K_s$ photometry
at a level of a few milli-magnitudes relative to 2MASS standards.
   }
   {}

         \keywords{
Astrometry --
Stars: distances --
Stars:  kinematics  --
Stars: low-mass, brown dwarfs --
subdwarfs --
solar neighbourhood
}

   \maketitle
%

\section{Introduction}

Subdwarfs were originally defined by Kuiper~(\cite{kuiper39})
as stars with spectral types A-K lying ``not over 2-3 mag
below the main sequence'' in optical colour-magnitude diagrams. 
They are low-metallicity stars which have typically large 
space velocities. They are
local representatives of the Galactic thick disk and halo populations
and show up preferentially in high proper motion surveys. However,
compared to thin disk stars with solar metallicities, subdwarfs
are a rare species. In the well-investigated close
solar neighbourhood ($d$$<$10~pc) there are only 4 subdwarfs
among the total of 348 objects
with accurately measured trigonometric parallaxes (Henry et al.~\cite{henry06}).

The coolest subdwarfs with late-K and M spectral types have been
described about ten years ago by Gizis~(\cite{gizis97}). He developed
a corresponding spectroscopic classification scheme separating 
normal M dwarfs with solar metallicities ($[M/H]\approx$0.0) from
M subdwarfs (sdM) with metallicities $[M/H]\approx$$-$1.2 and 
extreme subdwarfs (esdM) with even lower metallicities ($[M/H]\approx$$-$2.0).
The latest-type subdwarf classified by Gizis~(\cite{gizis97}) is the
high proper motion star LHS~377 with a spectral type of sdM7.

The research of very low-mass stars and brown dwarfs has 
experienced an enormous progress for the last decade, mainly
thanks to new deep optical and near-infrared (NIR) all-sky 
surveys, which also enable and support new high proper motion searches.
Many objects even cooler than M dwarfs have been discovered,
the new spectral types L and T have been invented to describe them
(see Kirkpatrick~\cite{kirkpatrick05} and references therein),
and their current census has reached nearly 700 objects (Gelino et al.~\cite{gelino08}).  
A question of particular interest is the formation of low-mass
stars and brown dwarfs in the low-metallicity regime. Relicts
of the early Galaxy and the first generations of star formation
may be detectable among the faintest objects in the local Galactic halo
population.
Interestingly, the new sky surveys have also revealed a new population
of ultracool subdwarfs (hereafter UCSDs), 
i.e. metal-deficient stars and brown dwarfs of spectral types later 
than sdM7 extending into the new L and possibly T spectral
classes (L{\'e}pine et al. \cite{lepine03b}; \cite{lepine03c};
Scholz et al.~\cite{scholz04a}; \cite{scholz04b}; 
Sivarani et al.~\cite{sivarani04};
Burgasser et al.~\cite{burgasser02}; \cite{burgasser03a};
\cite{burgasser04}; \cite{burgasser07};
Burgasser \& Kirkpatrick~\cite{burgasser06};
L{\'e}pine \& Scholz~\cite{lepine08}).

Trigonometric
parallax measurements of these new UCSDs are essential to
determine their absolute brightnesses, effective temperatures and
space motions. The first (sub-)stellar UCSDs with accurate
distance estimates will serve as benchmark sources for our understanding
of this new population and for their detailed classification. A new
classification scheme for UCSDs is still under debate 
(Gizis \& Harvin \cite{gizis06}; Burgasser et al. \cite{burgasser07}).
Recently, L{\'e}pine et al.~(\cite{lepine07}) have revised
the Gizis~(\cite{gizis97}) scheme for M subdwarfs and introduced a third 
class, the so-called ultrasubdwarfs (usdM), for the lowest metallicities.
In contrast, Jao et al.~(\cite{jao08}) have considered novel
methods for assigning spectral types from K3 to M6 dwarfs, discrediting the 
previous subdwarf metallicity classes and
suggesting instead a more complex investigation of temperature, 
metallicity and gravity features. Such a three-dimensional scheme
(temperature/clouds, metallicity, gravity)
has also been proposed by Kirkpatrick~(\cite{kirkpatrick05}) for
late-M, L and T dwarfs. However, the numbers of well-investigated 
subdwarfs are still too small (see Burgasser et al. \cite{burgasser07}
for an overview) to fill the required grid of subtypes with benchmark 
sources.
%
\begin{table}
\caption{Targets with previously known proper motions, $JHK_s$ photometry 
and spectral types}
\label{prevdata}
\begin{tabular}{@{\extracolsep{-6pt}}lccccc}
\hline\hline
Name    & $J$   &$H$   &$K_s$ &SpType&Ref \\ 
        & \multispan{3}{\hfil (2MASS) \hfil}   & & \\
(1)           & (2)&(3)   &(4)  &(5)  &(6)  \\ %
\hline
2MASS~0532$+$8246         & 15.179&14.904&14.918&sdL7                & 5;5,11  \\ 
2MASS~0937$+$2931         & 14.648&14.703&15.267&d/sdT6              & 7;1,11  \\ 
SSSPM~1013$-$1356         & 14.621&14.382&14.398&sdM9.5              & 3;3,11  \\ 
SSSPM~1256$-$1408         & 14.011&13.618&13.444&                    & 12; -   \\ 
SDSS~1256$-$0224          & 16.099&15.792&15.439&sdL4:               & 12;4,11  \\ 
LSR~1425$+$7102           & 14.775&14.405&14.328&sdM8                & 12;8,11  \\ 
SSSPM~1444$-$2019         & 12.546&12.142&11.933&d/sdM9              & 10;10,11  \\ 
LSR~1610$-$0040           & 12.911&12.302&12.019&d/sdM7:$^{\dagger}$ & 12;2,11  \\ 
2MASS~1626$+$3925         & 14.435&14.533&14.466&sdL4                &  6;6,11  \\ 
LSR~2036$+$5059$^{\star}$ & 13.611&13.160&12.936&sdM7.5              & 12;9,11  \\ 
\hline
\end{tabular}

\smallskip

\footnotesize{References (col. 6, first index for proper motion, others for sp. type):\\
1 - Burgasser et al.~(\cite{burgasser02}); 2 - L{\'e}pine et al.~(\cite{lepine03c});
3 - Scholz et al.~(\cite{scholz04a}); 4 - Sivarani et al.~(\cite{sivarani04});
5 - Burgasser et al.~(\cite{burgasser03a}); 6 - Burgasser~(\cite{burgasser04});
7 - Vrba et al.~(\cite{vrba04}); 8 - L{\'e}pine et al.~(\cite{lepine03b});
9 - L{\'e}pine et al.~(\cite{lepine03a});
10 - Scholz et al.~(\cite{scholz04b}); 
11 - Burgasser et al.~(\cite{burgasser07});
12 - Scholz et al. (unpublished, preliminary proper motion solution based on available SSS and 2MASS data).\\
$^{\dagger}$ -  discovered by L{\'e}pine et al.~(\cite{lepine03c}) as the first 
possible L subdwarf. Cushing \& Vacca~(\cite{cushing06}) described it as a very peculiar 
object (M6p/sdM), and Dahn et al.~(\cite{dahn08}) recently found it to be an astrometric 
binary of the Galactic halo population consisting of a mildly metal-poor M dwarf and a
substellar companion. \\
$^{\star}$ - L{\'e}pine et al.~(\cite{lepine02}) listed an erroneous position and proper motion
for this object.

}
\end{table}

\section{Target selection}

In 2004, when we initiated our subdwarf parallax programme, there were
only nine UCSDs known, and all happened to be visible from a northern
telesope site like Calar Alto. Among them, there were five M-type objects
originally detected in optical high proper motion surveys by L{\'e}pine
and co-workers (LSR~1425$+$7102, LSR~1610$-$0040, LSR~2036$+$5059)
and Scholz and co-workers (SSSPM~1013$-$1356, SSSPM~1444$-$2019),
three L- and T-type objects originally detected in the NIR
Two Micron All Sky Survey (2MASS; Skrutskie et al.~\cite{skrutskie06})
by Burgasser and co-workers (2MASS~0532$+$8246, 2MASS~0937$+$2931,
2MASS~1626$+$3925), and one object detected in the spectroscopic
data base of the deep optical Sloan Digital Sky Survey (SDSS;
York et al.~\cite{york00}) by Sivarani et al.~(\cite{sivarani04}).
The 2MASS photometry, as well as 
spectral types from different sources are given in Table~\ref{prevdata}.
We have included one more object (SSSPM~1256$-$1408), also detected in
the high proper motion survey by Scholz et al. (unpublished) using the SuperCOSMOS
Sky Surveys (SSS) data (Hambly et al.~\cite{hambly01}), which is
still lacking a spectral type.
Its large optical-to-NIR ($R$$-$$J$=$+$4.8; $R$ from SSS) and small
NIR ($J$$-$$K_s$=$+$0.57) colour indices are however typical of late-M
UCSDs.

When we started our observations, only one of our targets (2MASS~0937$+$2931)
had a preliminary trigonometric parallax measurement by Vrba et al.~(\cite{vrba04}). 
Meanwhile, there are three more parallaxes, all published in 2008, for
2MASS~0532$+$8246 (Burgasser et al.~\cite{burgasser08}), 
LSR~1425$+$7102 and LSR~1610$-$0040 (Dahn et al.~\cite{dahn08}).

\section{Observations and data reduction}~\label{obs}

The observations have been made with the OMEGA2000 camera
at the 3.5m-telescope of the Centro Astron\'omico Hispano Alem\'an (CAHA)
at Calar Alto, Spain. 
OMEGA2000 is a prime-focus, near-infrared, wide-field camera that
uses a 2k$\times$2k HAWAII-2 focal plane array
with a sensitivity from the $z$ to the $K$ band.
The optics of the camera consists of
a cryogenic focal reducer providing a 15.4$\arcmin$ $\times$ 15.4$\arcmin$
field of view with a resolution
of 0.45$\arcsec$/pixel.
The astrometric observations were all obtained in the $H$-band.
In all cases, at each epoch, we took 16 individual exposures of 60 s
each (frames), with small offsets of a few arcseconds, thus totalling 16 minutes
exposure time per object and night.
The observations were taken between January 2005 and June 2008. The
maximum epoch difference per target ranges from 3.1 to 3.4 years, with
the number of useful epochs (nights) from 17 to 26. In addition,
one observation (i.e., 16 individual frames) in the $J$ and $K_s$ bands has been
taken for each target.

Object detection and centroiding was carried out by use of the SEXTRACTOR software
(Bertin \& Arnouts~\cite{bertin96}).
For the photometric reduction, standard stars are taken from 2MASS.
On average, 100 stars per field with an accuracy better than 0.1~mag in 2MASS
were used as a reference.
In a given photometric band, each of 16 frames has been reduced to the
2MASS photometric system, separately. As a rule, a linear fit was sufficient
for stars fainter than 9th mag. For each object in a field, the  final 
$J$-, $H$,-  and $K_s$-magnitudes were computed as averages of 16 values. 
The limiting magnitudes are slightly changing from field to field and they reach
$J$ = 19, $H$ = 18, $K_s$ = 17.5, at least.

For each target, the astrometric data reduction was performed in several steps.
At first, an appropriate ``reference'' frame was chosen. Each frame was reduced to the 
``reference'' frame using a classical 2nd-order polynomial fit.  As reference points for
the ``frame-to-frame'' reduction we use anonymous field stars with $H$ magnitudes between
14 and 16.5; the number of reference stars varied between 75 for target 2MASS 0937+29
and 1250 for LSR 2036+5059.  The
``reference'' frame was transformed to an intermediate equatorial reference
system defined by 2MASS stars in a given sky area. Again, a 2nd-order
plate fit was carried out. The number of reference stars from 2MASS varied between
85 and 1650, the latter in the field around LSR 2036+5059.

\begin{table*}
\caption{Absolute parallaxes ($\pi(abs)$), absolute proper motions ($\mu_{\alpha}\cos \delta$, $\mu_{\delta}$),
and infrared magnitudes  $(J, H, K_s)$
of ultracool Subdwarfs}
\label{tab:results}
\centering
\begin{tabular}{@{\extracolsep{-2pt}}lrrrrrrrrrrr}
\hline\hline
\multicolumn{1}{c}{Name}&\multicolumn{1}{c}{RA J2000.0}  &\multicolumn{1}{c}{Dec J2000.0}  &\multicolumn{1}{c}{$\pi(abs)$}
&\multicolumn{1}{c}{$\Delta_{\pi}$} &$\mu_{\alpha}\cos \delta$  &\multicolumn{1}{c}{$\mu_{\delta}$} &\multicolumn{1}{c}{$J$} 
&\multicolumn{1}{c}{$H$}  &\multicolumn{1}{c}{$K_s$}  &\multicolumn{1}{c}{ $M_{K_s}$}& $V_{t(LSR)}$\\
     &\multicolumn{1}{c}{[h]} &\multicolumn{1}{c}{[deg]} & [mas] & [mas]   &[mas/yr]&[mas/yr]&[mmag]&[mmag]&[mmag]&[mag]&[km/s]        \\
\hline

2MASS 0532+82 &  5.548452 & 82.779208 & 42.28   & -3.36   & 2039.46 & -1661.79& 15145& 14894& 14904 & 13.03    & 289\\
              &           &           &$\pm$1.76&$\pm$1.37&$\pm$1.52&$\pm$1.64&$\pm$8&$ \pm$5&$\pm$15&$\pm$0.09&$\pm$  12\\
2MASS 0937+29 &  9.626350 & 29.528189 &163.39   &-3.39    &  944.15 & -1319.78&14622 & 14677& 15407 & 16.47    &  53 \\
              &           &           &$\pm$1.76&$\pm$1.18&$\pm$1.24&$\pm$1.21&$\pm$4&$\pm$7&$\pm$14&$\pm$0.03 &$\pm$   1\\
SSSPM 1013-13 & 10.218708 &-13.939245 & 20.28   &-5.11    &   69.44 & -1028.93&14637 & 14372& 14303 & 10.84    & 241\\
              &           &           &$\pm$1.96&$\pm$1.24&$\pm$1.20&$\pm$1.33&$\pm$7&$\pm$5&$\pm$ 7&$\pm$0.21 &$\pm$  23\\
SSSPM 1256-14 & 12.937228 &-14.144533 & 18.76   &-0.38    & -741.11 & -1002.13&14040 & 13624& 13458 &  9.82    & 305\\
              &           &           &$\pm$1.85&$\pm$1.10&$\pm$1.40&$\pm$1.38&$\pm$4&$\pm$5&$\pm$ 5&$\pm$0.21 &$\pm$  31\\
SDSS 1256-02  & 12.943648 & -2.414587 & 11.10   &-0.43    & -512.09 &  -297.71&16157 & 16060& 16061 & 11.29    & 242\\
              &           &           &$\pm$2.88&$\pm$1.11&$\pm$1.90&$\pm$1.79&$\pm$13&$\pm$8&$\pm$22&$\pm$0.56&$\pm$  66\\
LSR 1425+7102 & 14.418059 & 71.035998 & 12.19   &-0.73    & -602.38 &  -177.71&14828 & 14412& 14245 &  9.68    & 240\\
              &           &           &$\pm$1.07&$\pm$0.67&$\pm$0.98&$\pm$0.99&$\pm$7&$\pm$8&$\pm$ 8&$\pm$0.19 &$\pm$  21\\
SSSPM 1444-20 & 14.738983 &-20.323730 & 61.67   &-2.41    &-2906.15 & -1963.12&12602 & 12149& 11952 & 10.90    & 261\\
              &           &           &$\pm$2.12&$\pm$1.48&$\pm$2.41&$\pm$2.71&$\pm$6&$\pm$4&$\pm$ 4&$\pm$0.07 &$\pm$   9\\
LSR 1610-0040 & 16.174711 & -0.681642 & 33.10   &-2.63    & -773.84 & -1231.58&12872 & 12304& 12004 &  9.60    & 205\\
              &           &           &$\pm$1.32&$\pm$0.95&$\pm$0.91&$\pm$0.88&$\pm$7&$\pm$2&$\pm$ 5&$\pm$0.09 &$\pm$   8\\
2MASS 1626+39 & 16.438927 & 39.422076 & 29.85   &-1.10    &-1374.14 &   238.01&14426 & 14464& 14464 & 11.84    & 219\\
              &           &           &$\pm$1.08&$\pm$0.48&$\pm$0.96&$\pm$0.87&$\pm$5&$\pm$9&$\pm$10&$\pm$0.08 &$\pm$   8\\
LSR 2036+5059 & 20.606002 & 51.001279 & 21.60   &-1.00    &  751.93 &  1252.22&13628 & 13232& 12969 &  9.64    & 311\\
              &           &           &$\pm$1.26&$\pm$1.13&$\pm$1.10&$\pm$1.31&$\pm$9&$\pm$5&$\pm$20&$\pm$0.13 &$\pm$  19\\
\hline\hline
\end{tabular}

\end{table*}

In this intermediate system, mean positions,
proper motions and parallax were obtained via 
a rigorous single least-squares fit to the 5 unknowns, coupling the observational equations
in R.A. and Dec. via the parallax factor.
Although mathematically correct, the solution may suffer from a
relatively short time baseline and a non-uniform distribution of observations.
Thus correlations may influence the results for proper motions and parallax.  
In order to check the robustness of the 
solutions, two additional least-squares procedures were carried out. The first of these
treated the equations in R.A. and Dec. separately, yielding two solutions for the parallax.
In this check, correlations between R.A. and Dec. are prohibited. Evidently, the formal accuracy of
the parallax is always better from R.A. than from Dec. Therefore, a weighted mean parallax was computed.
The second checking approach is an iterative one. At first, only proper motions are computed.
The solutions are taken to remove the proper motion effect from the observed image displacements.
From the residuals, the parallax (and mean position again) is determined from the coupled equations.
The iterations start with the removal of the parallax effect from the original observations, and
determining proper motions from the latter. The procedure converged after 2 to 3 iterations. 
In this check, correlations between proper motions and parallax are prohibited.
For all ten targets, the three solutions for the parallaxes differed by less than 40\% of their
combined mean error. Therefore, we consider our rigorous solution as reliable, stable and robust.

Because of the deep observations and
the large field of 15.4$\arcmin$ $\times$ 15.4$\arcmin$, an appropriate number of galaxies
was found in each field, which were used to reduce the relative 
parallaxes and relative proper motions to absolute ones. 
Putting to zero the apparent parallaxes and proper motions of the galaxies yields the
corrections converting relative parallaxes and relative proper motions 
to absolute ones for all other objects in the field. 
The images of all galaxies in the fields were visually inspected to select sufficiently 
compact, and well measured reference objects. The number of useful
galaxies varied between 12 in the case of 2MASS 0532+82 to 105 in the case of
SSSPM 1256-14.

\section{Results and discussion}~\label{res}

Table 2 compiles the astrometric and photometric results for the 10 targets. 
The first column contains the target's name. Columns 2 and 3 give the mean (barycentric) position
(R.A. and Dec.) of the target for equinox and epoch J2000.0. Columns 4 and 5 present
the absolute parallaxes $\pi(abs)$ and the applied corrections $\Delta_{\pi}$ converting
the relative parallaxes to absolute ones as $\pi(abs) = \pi(rel) - \Delta_{\pi}$, while
columns 6 and 7 give the derived absolute proper motions of the targets. Columns
8, 9, 10 summarize the results of our photometric measurements. Column 11 gives the 
absolute magnitude $M_{K_s}$ computed from the corresponding trigonometric parallax and $K_s$,
whereas the last column lists the tangential velocities corrected for solar motion. 

   \begin{figure}
   \centering
   \includegraphics[bb=44 75 567 793,angle=-90,width=9.250cm,clip]{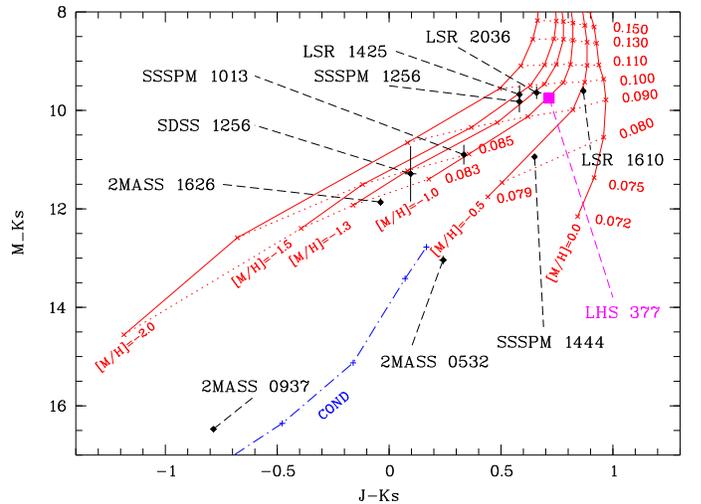}
      \caption{Colour-absolute magnitude diagram $M_{K_s}$ versus $J-K_s$ containing our
      10 targets and the sdM7 LHS~377 (filled square), a late type subdwarf with an
      already measured parallax (Monet et al.~\cite{monet92}; Gizis~\cite{gizis97}), plotted here for
      completeness. The error bars correspond to 1$\sigma$ errors. The full (red)
      lines represent the absolute magnitudes and colours derived from the
      theoretical models by Baraffe et al.~(\cite{baraffe97}) for
10~Gyr old objects with $[M/H]$ from $-$2.0 to $-$1.0,
and by Baraffe et al.~(\cite{baraffe98}) for 5~Gyr old
objects with $[M/H]$ from $-$0.5 to zero. The model data points were converted from
the CTI photometric system to the 2MASS system using the Carpenter~(\cite{carpenter01})
transformations. Dotted lines are the lines of equal mass.
The pluses connected by the dashed-dotted line show COND
model points for 5~Gyr old objects with less than 0.075 solar masses
(Baraffe et al.~\cite{baraffe03}).
              }
         \label{cmd}
   \end{figure}
%
Four of our targets have trigonometric parallaxes already published. These are
2MASS 0532+82 with $\pi = 37.5 \pm 1.7$~mas ($M_{K_s} = 12.79 \pm 0.18$~mag) by Burgasser et al.(~\cite{burgasser08}),  
2MASS 0937+29 with $\pi = 162.84 \pm 3.88$~mas ($M_{K_s} = 16.46 \pm 0.14$~mag) by Vrba et al.~(\cite{vrba04}),
LSR 1425+7102 and LSR 1610-0040 with, respectively, $\pi = 13.37 \pm 0.51$~mas ($M_{K_s} = 9.97 \pm 0.13$~mag) 
and $\pi = 31.02 \pm 0.26$~mas ($M_{K_s} = 9.48 \pm 0.04$~mag) both by Dahn et al.~(\cite{dahn08}).
All these parallaxes are based on 
observations with the 1.55~m Strand Astrometric Reflector at the
USNO Flagstaff Station. The number of reference stars used varied between 6 and 16. 
Photometric data were taken mainly from the 2MASS Catalog. In spite of the different
observation and reduction techniques, the published parallaxes coincide reasonably
well with our results, i.e. the differences
between two corresponding values are smaller than 2 times their mean errors. The parallax solutions 
from the observations with the 1.55~m Strand Astrometric Reflector take advantage of a better scale 
($\approx 0.35\arcsec$/pixel versus $\approx 0.45\arcsec$/pixel), and, for LSR 1425+7102 and LSR 1610-0040,
of considerably larger ranges of epochs and numbers of nights. On the other hand, we benefit from a better
SNR for the fainter targets, especially for photometry where the 2MASS Catalog is almost at its
limit. Moreover, a relatively wide and deep field allows to use a large number of reference stars and, 
for the first time, to derive parallaxes directly with
respect to galaxies and, consequently, to reduce a possible systematic bias in the parallax determination.

From the tangential velocities we infer that all objects 
except 2MASS 0937+29 exhibit halo kinematics.
Postponing the discussion of the latter, we find that the remaining targets
split into two groups in the CMD of Fig. \ref{cmd}. The first group is related to 
metallicity $[M/H] = -0.5$ in the Baraffe et al.~(\cite{baraffe97}) isochrones
for metal-poor low-mass stars, and is occupied by the objects
LSR 1610-0040, SSSPM 1444-20 and 2MASS 0532+82.
Ordered by luminosity they were classified as
d/sdM7, d/sdM9 and sdL7 in Burgasser et al.~(\cite{burgasser07}), respectively. For the first two objects,
this is consistent
with the Gizis~(\cite{gizis97}) classification scheme that rates dwarfs ``d'' with $[M/H]\approx 0$, and
subdwarfs ``sd'' with $[M/H] \approx -1.2$. Note that 2MASS 0532+82, extensively discussed by
Burgasser et al.~(\cite{burgasser08}) fits perfectly to the locus of $[M/H] = -0.5$, and
therefore is not an real subdwarf according to the Gizis~(\cite{gizis97}) classification. 
Its moderately low metallicity has also been suggested by
Scholz et al.~(\cite{scholz04b}) based on comparison with model colours ($I$$-$$J$
and $J$$-$$K$). Therefore, we think these three objects can be used
as benchmarks for  the type ``d/sd'' with 
$[M/H] = -0.5$ from M7 to L7.

Of the remaining 6 objects only one has a previously determined trigometric
parallax, LSR 1425+7102, measured by Dahn et al.~(\cite{dahn08}), and classified as
sdM8 by Burgasser et al.~(\cite{burgasser07}). The sdM7 LHS~377, which was not on our target list,
is a close neighbour to LSR 1425+7102 in the CMD.
 All our 6 objects populate the area between
$[M/H] = -1.0$ and $[M/H] = -1.5$ in the Baraffe et al.~(\cite{baraffe97}) isochrones. All
are classified as ``sd'' by Burgasser et al.~(\cite{burgasser07}) except our newly
detected object SSSPM 1256-14, which we
would classify as sdM8 based on its position in the CMD. The coolest object, the sdL4
2MASS 1626+39 has $M/M_{\odot} = 0.083$ and $T_{eff}$ = 2300~K when compared with
the Baraffe et al.~(\cite{baraffe97}) isochrones for $[M/H] = -1.0$. 
Our faintest (by apparent magnitude) target SDSS 1256-02 had the $K_s$ magnitude
in 2MASS given with a problem flag. It's 2MASS colour
of $J$$-$$K_s$ = 0.66 changes into 0.09 according to our photometry, and
hence its metallicity in the Baraffe et al.~(\cite{baraffe97}) models changes from $-$0.5 to $-$1.3.

Our 6 targets as well as LHS~377 serve as benchmarks for the subdwarf population between M7 (LHS~377)
and L4 (2MASS 1626+39) in the metallicity range between $[M/H] = -1.0$ and $ -1.5$.
Baraffe et al.~(\cite{baraffe03}) recently published new evolutionary models
for the coolest brown dwarfs (T dwarfs), which they refer to as the COND models.
In these models, dust opacity in
the radiative transfer equation is neglected. The COND isochrone is shown in Fig. \ref{cmd}
only for objects with $M/M_\odot$ $<$ 0.075. For masses up to $M/M_\odot$ = 0.08 it formally coincides with the $[M/H] = -0.5$
isochrone.
So, our targets from the $[M/H] = -0.5$ group
have loci close to the
COND isochrone, but the targets from the $[M/H] = -1.3$ class lie significantly above
this isochrone. 

2MASS 0937+29 may be the faintest object along the
extrapolated $[M/H] = -0.5$ isochrone in Fig.~\ref{cmd}. It was also
characterised as slightly metal-poor ($-0.4$$<$$[M/H]$$<$$-0.1$) by
Burgasser et al.~(\cite{burgasser03b}; \cite{burgasser06b}) using
spectral model comparisons. Compared to the 2MASS colour, our new
photometry lead to a bluer
$J-K_s$, which is again supporting a sub-solar metallicity if we
compare its location in Fig.~\ref{cmd} with the COND model points.
Its relatively low tangential velocity $V_{t(LSR)}$ of about 50~km/s
does not, however, reject a higher spatial velocity of 2MASS 0937+29
with respect to the local standard of rest (LSR). The presently unknown $V_{rad}$ of
2MASS 0937+29, contributes to its space velocity components as
 $(U,\,V,\,W)_{LSR} = (38 - 0.64V_{rad},\,-30 - 0.21V_{rad},\,+0.74V_{rad})$.
Therefore, as long as the radial velocity of 2MASS 0937+29 is unknown, one cannot exclude the
possibility that 2MASS 0937+29 is a member of the thick disk or even of the
halo population.
\\

In summary, we have measured infrared trigonometric parallaxes of ten ultracool
subdwarf, for 6 of which for the first time. The absolute parallaxes have referred
to galaxies directly, also for the first time.
Compared to theoretical models, 4 stars
have moderately low metallicity, $[M/H] \approx -0.5$, whereas 6 are consistent with
$[M/H]$ between $-1.0$ and $-1.5$. Nine out of ten definitely show halo kinematics from
their tangential velocities, while 2MASS 0937+29 needs a large radial velocity to be
kinematically excluded as a member of the disk.


\end{document}